\documentclass[12pt]{article}

\usepackage[totalheight = 23cm, totalwidth = 17cm]{geometry}
\usepackage{amssymb,amsmath,amsfonts,amsbsy,graphicx,bm}

\usepackage{color,cancel,ulem}

\newcommand{\mpl}{m_{\rm Pl}}

\newcommand{\dis}[1]{\begin{equation}\begin{split}#1\end{split}\end{equation}}

\begin{document}

\begin{titlepage}

\begin{center}

{\LARGE \bf 
Eternal inflation in light of Wheeler-DeWitt equation
}

\vskip 1.0cm

{\large
Min-Seok Seo$^{a}$ 
}

\vskip 0.5cm

{\it
$^{a}$Department of Physics Education, Korea National University of Education,
\\ 
Cheongju 28173, Republic of Korea
}

\vskip 1.2cm

\end{center}

\begin{abstract}

 The Wheeler-DeWitt equation provides the probability distribution for the curvature perturbation, the gauge invariant quantum fluctuation of the inflaton.
 From this, we can find a tower of power spectra which is not found in a perturbative approach.
 Since the power spectrum for the modes that cross the horizon contributes to the uncertainty in the classical inflaton displacement, we obtain new  conditions for eternal inflation.
 In the presence of the patch in the higher excitations, the bound on the slow-roll parameter allowing eternal inflation is  given by at most $\epsilon \lesssim (2n+1)(H/m_{\rm Pl})^2$ with $n$ integer indicating the quantum number labelling the excitation.
 For large $n$, the bound on $\epsilon$ is relaxed such that eternal inflation can take place with even larger value of $\epsilon$.
 While the second law of thermodynamics implies that $n=0$ state is preferred, we cannot ignore such large $n$ effect since the nonlinear interaction inducing transitions to the $n=0$ state is suppressed.

\end{abstract}

\end{titlepage}

\newpage

\section{Introduction}

  Whereas our understanding on quantum gravity still remains incomplete, we expect that the quantum nature of gravity plays the crucial role in the early universe.
  For example, primordial quantum fluctuation is believed to be the origin of large scale inhomogeneities as we observe from cosmic microwave background (CMB) radiation \cite{Mukhanov:1981xt, Mukhanov:1990me}.
 In the inflationary cosmology \cite{Guth:1980zm, Linde:1981mu, Albrecht:1982wi}, quantum  effects which are not contained in general relativity may give rise to `eternal inflation' \cite{Steinhardt:1983, Vilenkin:1983xq, Linde:1986fc, Linde:1986fd, Goncharov:1987ir} (for a review, see, e.g., \cite{Guth:2007ng}) by preventing the decrease in the vacuum energy density given by $3\mpl^2 H^2$.
 Since eternal inflation takes place only when the change in horizon size $H^{-1}$   is sufficiently tiny, investigating whether there is a physical reason in quantum gravity to forbid eternal inflation has a close connection to the stability of de Sitter (dS) spacetime.
 If quantum gravity excludes the small slow-roll parameter $\epsilon=-\dot{H}/H^2$ as the dS swampland conjecture claims \cite{Obied:2018sgi}, eternal inflation is not allowed as well  \cite{Matsui:2018bsy, Dimopoulos:2018upl}.
 However, the conjecture has been refined \cite{Ooguri:2018wrx}, 
 \footnote{While the original dS swampland conjecture considers the condition on the slope and the curvature of the potential, it eventually states the condition on the geometric quantity, the Hubble parameter $H$ \cite{Seo:2018abc}.
 For discussion on the thermodynamic aspect of the conjecture, see, e.g., \cite{Seo:2019mfk}.
 For discussion on the conjecture in light of the wavefunction of  universe along the line similar to this article can be found in, e.g., \cite{Matsui:2020tyd}.
}
  under which $\epsilon$ may be small for sufficiently long enough time \cite{Seo:2019wsh, Cai:2019dzj} provided   $\mpl^2 \nabla^2V/V\sim -{\cal O}(1)$ is satisfied \cite{Garg:2018reu, Andriot:2018mav}. 
  Then eternal inflation may be consistent with the refined conjecture \cite{Kinney:2018kew} (see also \cite{Rudelius:2019cfh} for a connection to the refined conjecture with various examples).
  Nevertheless, other quantum gravity mechanism such as the backreaction coming from Hawking radiation can destabilize dS such that eternal inflation is difficult to occur \cite{Markkanen:2016jhg, Markkanen:2016vrp}.

In order to quantify the condition for eternal inflation, we need to investigate a size of the quantum fluctuation of inflaton.
For this purpose, we note that the geometry of spacetime during inflation is quasi-dS, in which the time translation invariance is spontaneously broken.
Then the quantum fluctuation of the scale factor, or equivalently, time combines with that of the inflaton to become physical, which we will call the curvature perturbation \cite{Cheung:2007st, Weinberg:2008hq} (see also \cite{Prokopec:2010be, Gong:2016qpq} for discussion in terms of the path integral).
 The curvature perturbation has a remarkable property that as it crosses the horizon, i.e., its wavelength is stretched beyond the horizon, the quantum effects from the non-commutativity are suppressed.
 As a result, the curvature perturbation begins to behave like the classical long wavelength fluctuation \cite{Guth:1985ya, Albrecht:1992kf}, which can be interpreted  as a generation of the uncertainty in the classical displacement of inflaton.
 More precisely, as time goes on, the curvature perturbation of wavenumber satisfying $k=aH$ begins to cross the horizon in sequence, then the uncertainty in the inflaton displacement is accumulated as given by  $(H/2\pi)^2 H\Delta t$. 
After the Hubble time $H^{-1}$, the uncertainty in the inflaton displacement is increased by $(H/2\pi)^2$.
 At the same time, a single causal patch exponentially expands and becomes $e^3 \simeq 20$ causal patches.
If the size of the inflaton displacement caused by  quantum fluctuations exceeds the classical variance of the inflaton $\dot{\varphi}  H^{-1}$ in at least one of these patches, the inflaton in this patch fluctuates up inducing  `eternal' exponential expansion of the patch.
 By ignoring nonlinear interaction and postulating that the quantum fluctuation forms a stochastic Gaussian noise, the Gaussian probability distribution for the inflaton displacement can be obtained by solving the Fokker-Planck equation  \cite{Starobinsky:1986fx} which describes the random walk of the inflaton trajectory.
 Then we learn that  for eternal inflation to take place, $\epsilon$ is much smaller than $H^2/\mpl^2$.

 The uncertainty in the inflaton displacement discussed above is obtained from the power spectrum, the two point correlator of the curvature perturbation in the cosmological perturbation theory.
 Meanwhile, the geometrodynamical approach (for a review, see, e.g., \cite{Kiefer:2004gr})  provides  the probability distribution for the curvature perturbation from which we can read off the power spectrum as well.
 As can be found in Arnowitt-Deser-Misner (ADM) formlism \cite{Arnowitt:1962hi}, a time translation generator, the Hamiltonian involving both gravity and matter parts  is a constraint hence annihilates the physical states.
 This condition is written as the Wheeler-DeWitt (WDW) equation \cite{Wheeler:1964, DeWitt:1967yk}, the  equation for the `wavefunction of universe'.
 While time is not explicit in WDW equation, we can define time in the semiclassical limit by specifying the clock field  that parametrizes `how matter follows gravity' \cite{Banks:1984cw} (see also \cite{Brout:1987tq, Brout:1987ya, Brout:1988ku}).
 Then  the WDW equation has the form of Schr\"odinger equation.
 In the minisuperspace in which the degrees of freedom are restricted to the scale factor and the inflaton, we expect that the probability distribution for the curvature perturbation is obtained by solving it \cite{Halliwell:1984eu}.
  \footnote{ For recent studies on the inflationary cosmology in the context of the WDW equation can be found in, e.g., \cite{Kamenshchik:2013msa, Kamenshchik:2014kpa, Kamenshchik:2016mwj}.
  For a discussion on the dark energy, in which dS background is considered as well, in the context of WDW equation can be found in, e.g., \cite{Alon:2020yjq}.}
Indeed, when the nonlinear interaction is negligibly small, the WDW equation can be regarded as the Schr\"odinger equation for the harmonic oscillator with time dependent mass and frequency \cite{Kiefer:2011cc} (see also \cite{Kiefer:1990pt, Bertoni:1996ms, Kamenshchik:2017kfs}).

 Then it turns out that one of solutions which corresponds to the ground state gives the Gaussian probability distribution for the curvature perturbation, and the  uncertainty in this case is consistent with the power spectrum in the cosmological perturbation theory, as given by $(H/2\pi)^2$  \cite{Kiefer:2011cc, Brizuela:2019jzv}.
Moreover, the WDW equation is also solved by a tower of wavefunctions, which correspond to the higher excitations of the oscillator  \cite{Brizuela:2019jzv}.
The salient feature of these additional solutions is that as the excitation number gets larger, the probability is no longer peaked at the classical trajectory and  the uncertainty gets larger.
 This can be understood from an analogy with the simple  harmonic oscillator, in which the probability to detect the particle is maximized at the turning point, where the velocity of the oscillator vanishes.
Such a different behavior of probability distribution provides another type of eternal inflation condition when some patches in the universe have large excitation numbers.
 As the uncertainty in the inflaton displacement gets larger, we expect that the condition for eternal inflation can be relaxed.
 Our study confirms this by showing that first,  Gaussian is a good description of the probability distribution  for the {\it inflaton displacement} even if the {\it curvature perturbation} has a different probability distribution, and second, the condition for eternal inflation is given by at most $\epsilon \lesssim (2n+1)(H/\mpl)^2$, where the excitation number $n$ is an integer.
 Hence, in the presence of the causal patch having large $n$,  the  larger value of $\epsilon$  allows eternal inflation.  
 On the other hand, since the nonlinear interaction induces the transition between different excitations,  we may ask if there is a direction toward some specific state as a result of it. 
 As the number of patches increases exponentially, the complete calculation of the transition probability is quite nontrivial.
 Instead we may argue that the `ground state' is preferred by  the second law of thermodynamics, following \cite{Bousso:2006ge} (see also \cite{ArkaniHamed:2007ky, Wang:2019eym}).
 Of course, since the nonlinear interaction inducing the transition is suppressed by $\epsilon$ as well as $H/\mpl$, the transition probability is small, which implies that the effect from the states having large excitation number is not ignorable.

 This article is organized as follows.
 After a brief review on the WDW equation and settlement of the notation in section \ref{sec:WDW}, we move onto  section \ref{sec:WDWSol} to present the solution to WDW equation for   quasi-dS background, which has been studied in \cite{Brizuela:2019jzv}.
 While we  concentrate on the uncertainty in the curvature perturbation for the $k\gg a(t) |m|$ mode, we also find  that the uncertainty in the homogeneous limit, i.e., that for the $k \ll a(t)|m|$ mode also has interesting properties. 
 We postpone discussing this issue until appendix \ref{app:IRuncert} since it is out of the mainstream of our discussion.
  Solutions to the WDW equation are used in section \ref{sec:eternal} to give the condition for eternal inflation, which shows that for the large excitation number the bound on $\epsilon$ for eternal inflation is relaxed.
  We also discuss the implication of the new bound as well as effects from nonlinear interactions.
  After arguing that the second law of thermodynamics prefers the ground state, we conclude.

\section{WDW equation for FRW universe}
\label{sec:WDW}

  In ADM formalisn \cite{Arnowitt:1962hi}, the metric is decomposed in terms of a lapse $N$, a shift $N_i$, and a spatial metric $\gamma_{ij}$, 
\dis{ds^2=-N^2 dt^2+\gamma_{ij}(N^i dt+d x^i)(N^j dt+d x^j),} 
 from which the action for gravity and matter field $\varphi$ is written as
 \dis{S=\int d^3 x \Big[\pi^{ij}\partial_t {\gamma}_{ij}+\pi_\varphi \partial_t {\varphi}- N {\cal H}-N_i {\cal H}^i\Big].}
 In the last two terms,  ${\cal H}$ and ${\cal H}^i$ generate time- and spatial reparametrizations respectively, under which physics is invariant by general covariance.
  Non-dynamical $N$ and $N^i$ play the role of  Lagrange multipliers that accompany constraints : varying the action  with respect to them gives the conditions ${\cal H}={\cal H}^i=0$.
 Quantum mechanically, the constraints read
 \dis{{\cal H} |\Psi\rangle =0,\quad\quad {\cal H}^i |\Psi\rangle =0.}
  In particular the condition on ${\cal H}$ is called the WDW equation, which has been regarded as an equation for the `wavefunction of universe' \cite{Wheeler:1964, DeWitt:1967yk}.
  
 Now we consider spacetime geometry  described by  Friedmann-Robertson-Walker (FRW) metric
 \dis{ds^2=-dt^2+a^2(t)\Big[\frac{dr^2}{1-K r^2}+r^2d\theta^2+r^2\sin^2\theta d\phi^2\Big].}
 We note that by normalization leading to $K=0$ or $\pm1$ a scale factor $a(t)$ has a dimension $-1$ while $r$ is dimensionless. 
Then the action is written as
 \dis{S=\frac12\int d^4 x \Big[\mpl^2[-a\dot{a}^2+K a]+ a^3\Big[\frac12\dot{\varphi}^2-\frac12\frac{(\partial_i \varphi)^2}{a^2}-V(\varphi)\Big]\Big],}
 where $m_{\rm Pl}^2=(8 \pi G)^{-1}$, and if the matter potential is flat, i.e., $V(\varphi)=V_0$, it gives the cosmological constant $\Lambda=V_0/m_{\rm pl}^2$.
  From the canonical momenta,
  \dis{\pi_a=-6 m_{\rm Pl}^2a \dot{a},\quad\quad \pi_\varphi= a^3 \dot{\varphi}.\label{Eq:momenta}}
  the Hamiltonian density is given by
  \dis{{\cal H}&=\pi_a \dot{a}+\pi_\varphi \dot{\varphi}-{\cal L}
 \\
 &=\frac{1}{2a^3}\Big[-\frac{1}{6m_{\rm Pl}^2}a^2\pi_a^2+\pi_\varphi^2\Big]+ a^3\Big[\frac12\frac{(\partial_i\varphi)^2}{a^2}+V(\varphi)-2K\frac{m_{\rm Pl}^2}{a^2}\Big].\label{Eq:Hamiltonian}}
Classically, the constraint ${\cal H}=0$ is in fact nothing more than the Friedmann equation :  putting \eqref{Eq:momenta} into \eqref{Eq:Hamiltonian} with the homogeneity assumption $\partial_i \varphi=0$ we obtain
 \dis{\Big(\frac{\dot{a}}{a}\Big)^2=\frac{1}{3m_{\rm Pl}^2}\Big[\frac12\dot{\varphi}^2+V(\varphi)\Big]-\frac{K}{a^2}.}
 From now on we consider the flat FRW spacetime only, so we set $K=0$.
 
 In the field basis, the canonical momenta in the WDW equation, ${\cal H}|\Psi\rangle=0$ is represented by the functional derivatives.
 Since the  momentum part,
   \dis{\frac12\Big[-\frac{1}{6 m_{\rm Pl}^2 a}\pi_a^2 +\frac{1}{a^3}\pi_\varphi^2\Big]\equiv \frac12 G^{AB}\pi_A \pi_B,}
    is written with respect to the field space metric $G_{AB}={\rm diag.}(-6 m_{\rm Pl}^2 a, a^3)$,
  we replace the  momentum term by `Laplacian' as 
  \dis{G^{AB} \pi_A \pi_B \to -\frac{1}{\sqrt{-G}}\partial_A(\sqrt{-G} G^{AB}\partial_B)&=-\frac{1}{a^3}\Big[-\frac{1}{6m_{\rm Pl}^2}a\frac{\partial}{\partial a}\Big(a\frac{\partial}{\partial a}\Big)+\frac{\partial^2}{\partial\varphi^2}\Big]
  \\
  &=\frac{1}{a^3}\Big[\frac{1}{6m_{\rm Pl}^2}\frac{\partial^2}{\partial N_e^2}-\frac{\partial^2}{\partial\varphi^2}\Big],}
 where $N_e$ being the number of $e$-folds such that the WDW equation reads
  \dis{\frac{e^{-3N_e}}{2 a_0^3}\Big[\Big[\frac{1}{6m_{\rm Pl}^2}\frac{\partial^2}{\partial N_e^2}-\frac{\partial^2}{\partial\varphi^2}\Big]+ a_0^4e^{4N_e}(\partial_i\varphi)^2+ 2 a_0^6 e^{6N_e}V(\varphi)\Big]\Psi(N_e, \varphi)=0.\label{Eq:WDWfinal}}

\section{Solution to the WDW equation as a probability density}
\label{sec:WDWSol}

\subsection{WDW equation for the curvature perturbation}

 In the WDW equation \eqref{Eq:WDWfinal}, time does not appear as an explicit parameter.
 Instead the scale factor, or equivalently, the number of $e$-folds $N_e$ can play the role of the `clock field'.
 It  defines time in the semiclassical limit, from which the WDW equation has the form of the Schr\"odinger equation.
 Since we are interested in the probability distribution for the curvature perturbation, we also need to rewrite \eqref{Eq:WDWfinal} by including quantum fluctuations of the inflaton $\varphi$ and $N_e$ \cite{Halliwell:1984eu}.
 These quantum fluctuations are not independent but form the gauge invariant combination,   the Mukhanov-Sasaki variable \cite{Mukhanov:1985rz, Sasaki:1986hm}.
 It corresponds to the  scalar field representing the curvature perturbation,
 \footnote{In fact, time as well as time derivative here is defined after the clock field is specified as shown in the following discussion.
 Nevertheless, we use time derivative in advance since we are working in the semiclassical limit and also expect that the gauge invariant (hence physically meaningful) quantity which becomes the Mukhanov-Sasaki variable in the semiclassical limit can be defined.
 } 
 \dis{\tilde{\varphi}=\delta\varphi-\frac{\dot{\varphi}}{H}\delta N_e,}
 where $\dot{\varphi}$ is the time derivative of the classical trajectory of $\varphi$, which is connected to $\epsilon$ through
  \dis{\epsilon=\frac{\dot{{\varphi}}^2}{2 m_{\rm Pl}^2 H^2} \ll 1.\label{Eq:epsilon}}
 This shows that the time translation invariance which has been a part of dS isometry is spontaneously broken in the quasi-dS background, such that the fluctuation in time, or equivalently, that in $N_e$ becomes physical by `absorbing' the inflaton fluctuation.
 The dynamics of $\tilde{\varphi}$ is equivalent to that of the scalar field on the quasi-dS background, whereas detailed form of interactions reflects the nature of the trace part of the metric.
 Indeed, the quadratic action for $\tilde{\varphi}$ is simply that for the scalar field on quasi-dS background,
   \dis{S_2=\int d^4 x\frac{a^3}{2}[  \dot{\tilde \varphi}^2-\frac{1}{a^2}(\partial_i  {\tilde \varphi})^2-m^2  {\tilde \varphi}^2],}
   where the mass squared is given by $m^2=-3H \dot{\epsilon}/(2\epsilon)-\ddot{\epsilon}/(2\epsilon)+\dot{\epsilon}^2/(4\epsilon^2)$, roughly $H^2$ times  slow-roll parameter.
 \footnote{Typically, the curvature perturbation focuses on the fluctuation in $N_e$, by defining ${\cal R} =\delta N_e-\frac{H}{\dot{\varphi}}\delta\varphi$.
 Then the quadratic action for ${\cal R}$ is written as
 \dis{S_2=\int d^4 x \frac{a^3}{2}(2\epsilon \mpl^2)\Big[\dot{\cal R}^2-\frac{1}{a^2}(\partial_i {\cal R})^2\Big],}
 which reflects the spontaneous breaking of dS isometry in a more obvious way : $\epsilon$ is the order parameter for the dS isometry breaking, and the action looks like that for the Goldstone boson.
 In this article, we instead take an equivalent field $\tilde{\varphi}=\sqrt{2\epsilon}\mpl {\cal R}$ as a dynamical variable for the convenience of formal treatment.
\label{ft:Rphi}  }

 Then we can consider the WDW equation as a functional of classical trajectories $N_e$, $\varphi$ and the quantum fluctuation $\tilde{\varphi}$.
 In fact,   $N_e$ and $\varphi$ are not independent as they are connected through the  Friedmann equation, hence we take $N_e$ as the only classical variable in the WDW equation.
  In addition, in the slow roll approximation, $\pi_{ \varphi}= a^3 \dot{\varphi}$ is ignorable compared to the potential term since $\pi_{\varphi}^2 = a^6 \dot{\varphi}^2 \ll a^6 V =3 a^6  m_{\rm Pl}^2 H^2$ as implied by \eqref{Eq:epsilon} \cite{Kiefer:2011cc}.
 Then the WDW equation is written as 
   \dis{&\frac{e^{-3N_e}}{2 a_0^3}\Big[\Big[\frac{1}{6m_{\rm Pl}^2}\frac{\partial^2}{\partial N_e^2}-\frac{\partial^2}{\partial{\tilde \varphi}^2}\Big]+2 a_0^6 e^{6N_e}V(\varphi)
 +(-a_0^4e^{4N_e}\partial_i^2+m^2 a_0^6e^{6N_e})\tilde{\varphi}^2\Big]\Psi(\tilde{\varphi}, {N}_e)=0.\label{Eq:WDWreal}}
 
 In order to solve \eqref{Eq:WDWreal} in the semiclassical limit,  \cite{Kiefer:2011cc, Kiefer:1990pt} (see also \cite{Bertoni:1996ms, Kamenshchik:2017kfs} for a recent refinement concerning unitarity) suggested that we first consider the wavefunction in the WKB approximation form $\Psi \sim e^{iS}$ and then expand $S$ with respect to $m_{\rm Pl}$ as $S=m_{\rm Pl}^2 S_0 + S_1+\cdots$.
Here we summarize the results obtained in \cite{Kiefer:2011cc} below:
\begin{itemize}
\item At ${\cal O}(m_{\rm Pl}^4)$, we have
\dis{\Big(\frac{\partial S_0}{\partial \tilde{\varphi}}\Big)^2=0,\label{Eq:O4}}
hence $S_0$ is independent of quantum fluctuation $\tilde{\varphi}$.

\item At ${\cal O}(m_{\rm Pl}^2)$, 
\dis{-\frac{m_{\rm Pl}^2}{6}\Big(\frac{\partial S_0}{\partial N_e}\Big)^2 -m_{\rm Pl}^2\Big(i\frac{\partial^2 S_0}{\partial \tilde{\varphi}^2}-2 \frac{\partial S_0}{\partial  \tilde{\varphi}}\Big)+6a_0^6e^{6N_e} m_{\rm Pl}^2 H^2=0.}
The second parenthesis vanishes from \eqref{Eq:O4}.
This is equivalent to the Hamilton-Jacobi equation, the solution to which is given by
\dis{ S_0(N_e)=\pm 2 H a_0^3 e^{3N_e}. } 

\item At ${\cal O}(m_{\rm Pl}^0)$, 
\dis{\frac{e^{-3N_e}}{2a_0^3}\Big[&\frac{1}{6}\Big[i \frac{\partial^2 S_0}{\partial N_e^2}-2\frac{\partial S_0}{\partial N_e}\frac{\partial S_1}{\partial N_e}\Big]
- \Big[i \frac{\partial^2 S_1}{\partial  {\tilde \varphi}^2}- \Big(\frac{\partial S_1}{\partial  {\tilde \varphi}}\Big)^2\Big]
-(a_0^4 e^{4N_e}\partial_i^2-m^2 a_0^6e^{6N_e})  {\tilde \varphi}^2\Big]=0. \label{Eq:SchOri}}
We first note that  the second bracket can be interpreted as a part of 
\dis{ \frac{\partial^2 }{\partial  {\tilde \varphi}^2} e^{iS_1}=  \Big[i \frac{\partial^2 S_1}{\partial  {\tilde \varphi}^2}-\Big(\frac{\partial S_1}{\partial  {\tilde \varphi}}\Big)^2\Big] e^{iS_1}.}
Now, we can define `time' in the Banks' sense \cite{Banks:1984cw},
\dis{\frac{\partial}{\partial t}=-\frac{e^{-3N_e}}{6 a_0^3}\frac{\partial S_0}{\partial N_e}\frac{\partial}{\partial N_e}=H\frac{\partial}{\partial N_e}.}
Then  \eqref{Eq:SchOri} is equivalent to the Schr\"odinger equation for the wavefunction
\dis{\psi(N_e,  \tilde{\varphi})=\gamma(N_e)e^{iS_1(N_e,  \tilde{\varphi})},\quad\quad \gamma(N_e)=\sqrt{\frac{\partial S_0}{\partial N_e}},}
which is given by
\dis{i\frac{\partial}{\partial t}\psi 
=\frac{e^{-3N_e}}{2a_0^3} \Big[-   \frac{\partial^2 }{\partial {\tilde \varphi}^2} +(-a_0^4 e^{4N_e}\partial_i^2+m^2 a_0^6e^{6N_e})\Big]\psi \equiv H \psi.\label{Eq:Schr}}
\end{itemize}

The procedure above as studied in \cite{Kiefer:2011cc} shows that in the semiclassical limit we can define time from which the WDW equation is converted into the Schr\"odinger equation \eqref{Eq:Schr} for the curvature perturbation $\tilde{\varphi}$.
Then $|\psi|^2$ can be interpreted as a probability density for $\tilde{\varphi}$. 

\subsection{Solving the WDW equation}
\label{subsec:WDWSolcal}

 In this section, we solve the WDW equation in the Schr\"odinger equation form, \eqref{Eq:Schr} which is relevant to our discussion on eternal inflation.
 For this purpose, we consider massless ($m=0$) but inhomogeneous ($\partial_i \tilde{\varphi} \ne 0$) fluctuation.
  This is a good approximation for the extremely tiny slow-roll parameter, or the modes satisfying $k \gg a |m|$.
  From this, we obtain the power spectrum as the uncertainty in the probability distribution for $\tilde{\varphi}$ \cite{Brizuela:2019jzv}.
 As already known \cite{Vilenkin:1982wt, Linde:1982uu, Starobinsky:1982ee}, the power spectrum for the superhorizon mode contributes to the uncertainty in the classical inflaton displacement in an accumulative way.
On the other hand, when we are interested in the uncertainty in homogeneous field, we may consider the infrared mode $k\ll a|m|$, in which the curvature perturbation is almost homogeneous ($\partial_i \tilde{\varphi} \simeq 0$).
 Since this is not in the mainstream of our discussion concerning eternal inflation, we discuss some feature of this case in appendix \ref{app:IRuncert}.

In order to find the WDW equation in the case of $\partial_i \tilde{\varphi} \ne 0$,  we first consider the Fourier decomposition of $\tilde{\varphi}$ in momentum space,
\dis{\tilde{\varphi}(x)=\int \frac{d^3 k}{(2\pi)^{3/2}}e^{i \textbf{k} \cdot \textbf{x}}\tilde{\varphi}_{\textbf k}.}
Then the quadratic action in momentum space is written as
\dis{S_2=\int dt\frac{a^3}{2} \int\frac{d^3 k}{(2\pi)^3}[  \dot{\tilde \varphi}_{\textbf k}\dot{\tilde \varphi}_{-{\textbf k}}-\Big[\frac{k^2}{a^2}+m^2\Big] {\tilde \varphi}_{\textbf k}{\tilde \varphi}_{-{\textbf k}}  ].}
Treating the momentum as an index specifying different complex scalar fields (${\tilde \varphi}_{-{\textbf k}}={\tilde \varphi}_{\textbf k}^\dagger$ for real $\tilde{\varphi}(x)$), the Hamiltonian can be understood as a sum (more precisely, integration) of the Hamiltonian for each mode, 
\dis{{\cal H}_{\textbf k} = \frac{e^{-3N_e}}{2a_0^3} \Big[-  \frac{\partial^2 }{\partial {\tilde \varphi}_{\textbf k}\partial {\tilde \varphi}_{-{\textbf k}}} +( k^2 a_0^4 e^{4N_e}+m^2 a_0^6e^{6N_e}){\tilde \varphi}_{\textbf k} {\tilde \varphi}_{-{\textbf k}}\Big].\label{Eq:Hamkmode}}
Moreover, dividing $\tilde{\pi}_{-{\textbf k}}=-i\partial/\partial {\tilde \varphi}_{\textbf k}$ and ${\tilde \varphi}_{\textbf k}$ into real and imaginary parts, under  the ansatz for the wavefunction for ${\tilde \varphi}_{\textbf k}={\tilde \varphi}^R_{\textbf k}+i {\tilde \varphi}^I_{\textbf k}$,
 \dis{\psi(\{{\tilde \varphi}_{\textbf k}\})=\prod_\textbf{k} \psi^R_\textbf{k}({\tilde \varphi}^R_{\textbf k}) \psi^I_\textbf{k}({\tilde \varphi}^I_{\textbf k}),\label{Eq:kprod}}
 the Schr\"odinger equation for each mode is written as
 \dis{i\frac{\partial}{\partial t} \psi_{\textbf k}^A= \frac{e^{-3N_e}}{2a_0^3} \Big[-  \frac{\partial^2 }{\partial {\tilde \varphi}^A_{\textbf k}\partial {\tilde \varphi}^A_{{\textbf k}}} +( k^2 a_0^4 e^{4N_e}+m^2 a_0^6e^{6N_e}){\tilde \varphi}^A_{\textbf k} {\tilde \varphi}^A_{{\textbf k}}\Big]\psi_{\textbf k}^A,}
 with $A=R, I$.
 This is equivalent to the time-dependent harmonic oscillator, $H=\frac{1}{2M(t)}p^2+\frac12 M(t)\Omega(t)^2 q^2$ ($q: {\tilde \varphi}^A_{{\textbf k}}$, $p: {\tilde \pi}^A_{{\textbf k}}$), with
\dis{M(t)=  a_0^3 e^{3N_e},\quad\quad \Omega(t)^2=\frac{e^{-6N_e}}{a_0^6}[a_0^4 e^{4N_e}k^2+m^2 a_0^6e^{6N_e}].\label{Eq:HO}}

Solution to the Schr\"odinger equation for the time-dependent harmonic oscillator is already known \cite{Ji:1995zz, Pedrosa:1997}.
It makes use of the fact that, the Hamiltonian  which is at most quadratic in the variable and its canonical momentum has a constant Hermitian operator, the Lewis-Riesenfeld invariant $I$ \cite{Lewis:1968tm}.
Here, by `constant', we mean $I$ satisfies
\dis{\frac{dI}{dt}=\frac{\partial I}{\partial t}-i[H, I]=0.\label{Eq:invcond}}
The Lewis-Riesenfeld invariant for the time dependent harmonic oscillator is given by
\dis{I=\frac12\Big[\frac{q^2}{\rho_k(t)^2}+(\rho_k(t) p-M(t)\dot{\rho}_k(t) q)^2\Big].}
From the condition \eqref{Eq:invcond}, the time dependent function $\rho_k(t)$ satisfies
\dis{\ddot{\rho}_k+3\frac{\dot M}{M}\dot{\rho}_k+\Omega^2 \rho_k=\frac{1}{M^2 \rho_k^3}.}
 It has been found in \cite{Lewis:1968tm} that, the solution to the Schr\"odinger equation $i\dot{\psi}=H\psi$ is given by the superposition of the eigenstates $\phi_n$ of $I$ with the `gauge transformation' by the `Lewis phase' $\alpha_n$,  
 \dis{\psi(t)=\sum_n c_n e^{i\alpha_n (t)} \phi_n,}
 where the Lewis phase satisfies
 \dis{\frac{d \alpha_n}{d t}=\Big\langle \phi_n \Big| \Big[i\frac{\partial}{\partial t}-H\Big] \Big|\phi_n \Big\rangle.}
 What \cite{Ji:1995zz, Pedrosa:1997} pointed out is that, by unitary, or equivalently, canonical transformation, the eigenvalue problem of $I$ can be converted into that of the simple harmonic oscillator, in which both the mass and the frequency are time independent.
 As a result, the Lewis phase behaves like the phase $E_n t$ appearing in the time evolution of energy eigenstate,
 \dis{\alpha_n(t)=-\Big(n+\frac12\Big)\int^t\frac{dt'}{M(t')\rho_k(t')^2},}
 and the solution to the Schr\"odinger equation has the form similar to that of simple harmonic oscillator :
  \dis{\psi_{\textbf{k}, n}(t)=e^{i\alpha_n }\Big(\frac{1}{\sqrt{\pi}n! 2^n \rho_k}\Big)^{1/2}e^{\big[i\frac{M}{2}\frac{\dot\rho_k}{\rho_k}-\frac{1}{2\rho_k^2}\big]({\tilde{\varphi}_{\textbf k}^A})^2}H_n(\tilde{\varphi}_{\textbf k}^A/\rho_k).\label{Eq:solution}} 
  From the mass and the frequency  given by \eqref{Eq:HO}, the equation for $\rho_k$ in our case becomes
  \dis{\ddot{\rho}_k+3H\dot{\rho}_k+\Big[\frac{k^2}{a^2}+m^2\Big]\rho_k=\frac{1}{  a^6 \rho_k^3}.\label{Eq:rhomaster}}
  This $\rho_k$  plays the role of $1/\sqrt{M \Omega}$ and is closely connected to the uncertainty in the probability distribution as   $|\psi|^2$ is interpreted as a probability density.

 Now let us suppose $m=0$. 
 Then the differential equation for $\rho_k$,
  \dis{\ddot{\rho}_k+3H\dot{\rho}_k+\frac{k^2}{a_0^2}e^{-2 H t}\rho_k=\frac{e^{-6 H t}}{  a_0^6 \rho_k^3},}
  where we set $a=a_0 e^{H t}$, has a solution
  \dis{\rho_k=\frac{H}{k^{3/2}}\frac{\sqrt{k^2+(a H)^2}}{a H},\label{Eq:rhok}}
  as already obtained in \cite{Brizuela:2019jzv}.
  For sufficiently large $t$, $\rho_k \simeq H/k^{3/2}$ is almost constant in time.
  This result is remarkable as it provides another way to obtain the two-point correlator.
  Typically, the mode expansion for the massless scalar field on the perfect dS background is given by
  \dis{\tilde{\varphi}(x)=\int \frac{d^3 k}{(2\pi)^{3}}\frac{e^{i \textbf{k}\cdot \textbf{x}}}{\sqrt{2k} a}\Big[e^{-ik\tau}\Big(1-\frac{i}{k\tau}\Big)a_{\textbf k}+e^{ik\tau}\Big(1+\frac{i}{k\tau}\Big)a_{-{\textbf k}}^\dagger\Big],\label{Eq:modesxp}}
  with the conformal time  $\tau=-(aH)^{-1}$.
  Even though the curvature perturbation does not appear in   perfect dS, the expression above is a good approximation for the small slow-roll parameter, $\epsilon \ll 1$.
  The nature coming from spontaneous breaking of dS isometry is restored by taking  ${\cal R}=\tilde{\varphi}/(\sqrt{2\epsilon}\mpl)$ as the field variable  (see footnote \ref{ft:Rphi}).
  From \eqref{Eq:modesxp} the two-point correlator is given by
  \dis{\langle \tilde{\varphi}_{\textbf k}  \tilde{\varphi}_{{\textbf k}'}\rangle&=\frac{1}{2k a^2}\Big[1+\frac{1}{(k\tau)^2}\Big]\delta^3(\textbf{k}+\textbf{k}')=\frac{H^2}{2 k^3}\Big[1+\frac{k^2}{(aH)^2}\Big]\delta^3(\textbf{k}+\textbf{k}')
  \\
  &=\frac{\rho_k^2}{2}\delta^3(\textbf{k}+\textbf{k}'). }
  The connection between the two-point correlator and $\rho_k$ is clear from the wavefunction \eqref{Eq:solution} with $n=0$, which is Gaussian.
  To see this in detail, we first consider the `generating functional',
  \dis{Z=\int \prod_\textbf{k}d\tilde{\varphi}^R_\textbf{k} d\tilde{\varphi}^I_\textbf{k} P(\tilde{\varphi}_\textbf{k})e^{ \frac12[J^R_{\textbf{k}}\tilde{\varphi}^R_\textbf{k} + J^I_\textbf{k}\tilde{\varphi}^I_{\textbf{k}}]}
  =  \int \prod_\textbf{k}d\tilde{\varphi}^R_\textbf{k} d\tilde{\varphi}^I_\textbf{k}  \Big(\frac{1}{\pi \rho_k^2}\Big) e^{-\frac{(\tilde{\varphi}^R_{\textbf k})^2}{\rho_k^2}+\frac12 J^R_{\textbf{k}}\tilde{\varphi}^R_\textbf{k}}
    e^{-\frac{(\tilde{\varphi}^I_{\textbf k})^2}{\rho_k^2}+\frac12 J^I_{\textbf{k}}\tilde{\varphi}^I_\textbf{k}}.}
  The $1/2$ factor in the source term comes from the fact that $\tilde{\varphi}_{-\textbf{k}}$ is just $\tilde{\varphi}_{\textbf{k}}^\dagger$ hence $(\tilde{\varphi}^R_{\textbf{k}}, \tilde{\varphi}^I_{\textbf{k}})$ and  $(\tilde{\varphi}^R_{-\textbf{k}}, \tilde{\varphi}^I_{-\textbf{k}})=(\tilde{\varphi}^R_{\textbf{k}}, -\tilde{\varphi}^I_{\textbf{k}})$ are not independent variables.
 As a result, we can give the common source term $J_{-\textbf{k}}\tilde{\varphi}_\textbf{k}+J_\textbf{k} \tilde{\varphi}_{-\textbf{k}}$ for $\textbf{k}$ and $-\textbf{k}$ modes by  taking $(J^R_{-\textbf{k}}, J^I_{-\textbf{k}})=(J^R_{\textbf{k}}, -J^I_{\textbf{k}})$.
 The $1/2$ factor means that this common source term is divided in half into source terms for $\textbf{k}$ and $-\textbf{k}$ modes, respectively.
 Moreover, from $ J^R_{\textbf{k}}\tilde{\varphi}^R_\textbf{k} + J^I_\textbf{k}\tilde{\varphi}^I_{\textbf{k}}=J_{-\textbf{k}}\tilde{\varphi}_\textbf{k}+J_\textbf{k} \tilde{\varphi}_{-\textbf{k}}$, one finds that $J_{\pm\textbf{k}}=\frac12(J^R_\textbf{k}\pm iJ^I_\textbf{k})$.
 Then the generating functional is calculated to be 
 \dis{Z=\prod_\textbf{k}e^{\frac{\rho_k^2}{16}[(J_\textbf{k}^R)^2+(J_\textbf{k}^I)^2]}=e^{\int \frac{d^3 k}{(2\pi)^3} \frac{\rho_k^2}{4}J_\textbf{k}J_{-\textbf{k}}}.}
  From this,  the two point correlator can be understood as
  \dis{\langle \tilde{\varphi}_{\textbf k}  \tilde{\varphi}_{{\textbf k}'}\rangle &= \int \prod_\textbf{k}d\tilde{\varphi}_\textbf{k}\tilde{\varphi}_\textbf{k} \tilde{\varphi}_{\textbf{k}'} P(\tilde{\varphi}_\textbf{k})
  = \frac{\delta^2 Z}{ \delta J_{\textbf{k}} \delta J_{\textbf{k}'}} \Big|_{  J_{\textbf{k}} =J_{-\textbf{k}}=0}
  \\
&=\frac{\rho_k^2}{2}(2\pi)^3\delta^3(\textbf{k}+\textbf{k}').\label{Eq:2ptft}}

 Now, the expanding universe stretches the wavelength of the fluctuation such that after $t=(1/H)\log(k/H)$ at which $k=a(t) H$ is satisfied, $\tilde{\varphi}_{\textbf k}$  behaves like the fluctuation in the classical trajectory of the inflaton field value.
 As time goes on, the fluctuation in  $\tilde{\varphi}_{\textbf k}$ with the wavenumber satisfying $k=a(t)H$ begins to provide the accumulated uncertainty in the inflaton displacement.
 Let us suppose that the inflaton has a some specific classical value at $t_i$.
 Then the accumulated uncertainty generated during a time interval $t_f-t_i$ as seen at much later time is given by the integration of \eqref{Eq:2ptft} over momenta ${\textbf{k}}$ and ${\textbf{k}'}$,
 \dis{\langle {\varphi}(t_f)^2\rangle-\langle  {\varphi}(t_i)^2\rangle = \int \frac{d^3 k}{(2\pi)^3} \frac{d^3 k'}{(2\pi)^3} \langle \tilde{\varphi}_{\textbf k}  \tilde{\varphi}_{{\textbf k}'}\rangle = \int_{k_i}^{k_f} \frac{dk}{k}\Big(\frac{H}{2\pi}\Big)^2=
\Big(\frac{H}{2\pi}\Big)^2 \log\Big(\frac{k_f}{k_i}\Big).}
Here the upper- and the lower bounds of integration   satisfy $k_i=a(t_i)H$ and $k_f=a(t_f)H$, respectively, hence $dk/k$ is interpreted as $d(aH)/(aH)=d N_e$ and the ratio $\log(k_f/k_i)$ is identified with the variation $\Delta N_e = H (t_f-t_i)$.
 Then $(H/2\pi)^2$ becomes the uncertainty generated per unit $e$-fold.

For the higher excitations,   each of real and imaginary parts of ${\tilde \varphi}^A_{{\textbf k}}$ is governed by the probability distribution coming from \eqref{Eq:solution} with $n >0$.
Since it is equivalent to the probability distribution for the harmonic oscillator except for the replacement of $1/(M\Omega)$ by $\rho_k^2$, we expect that the uncertainty for the $\textbf{k}$ mode is given by $(n+\frac12)\rho_k^2$ as can be read off from the uncertainty for the $n$ th excitation of the harmonic oscillator.
This indeed is verified in the explicit calculation in \cite{Brizuela:2019jzv}.
 In terms of the harmonic oscillator, the uncertainty is interpreted as a half of the maximal displacement squared.
 That is, given energy $(n+\frac12)\Omega$, the `maximal displacement' from $\tilde{\varphi}_\textbf{k}=0$, $2(n+\frac12)\rho_k^2$ is obtained by equating the energy with $\frac12 M \Omega^2 \tilde{\varphi}_\textbf{k}^2$, and replacing $1/(M\Omega)$ by $\rho_k^2$.
 For large $n$, the probability is maximized at the maximal displacement.
 This is consistent with the classical interpretation of the simple harmonic oscillator that the velocity of a particle in the oscillator vanishes at the maximal displacement hence probability to detect the particle, $P\sim [2M[(n+\frac12)\Omega-\frac12M\Omega^2 \tilde{\varphi}_{\textbf k}^2]]^{-1/2}$  is maximized there.
 In our case, we just further need to replace $M\Omega$ by $1/\rho_k^2$, which gives $P \sim \rho_k[2(n+\frac12)-\tilde{\varphi}_{\textbf k}^2/\rho_k^2]^{-1/2}$. 
 If all the $\textbf{k}$ modes are in the same $n$th excitation, 
 integrating the uncertainty for the $\textbf{k}$ mode over momentum gives the accumulated uncertainty in the classical inflaton field value :
  \dis{\langle {\varphi}(t_f)^2\rangle-\langle  {\varphi}(t_i)^2\rangle =  (2n+1)\int_{k_i}^{k_f} \frac{dk}{k}\Big(\frac{H}{2\pi}\Big)^2=(2n+1)\Big(\frac{H}{2\pi}\Big)^2 \Delta N_e.} 
 Therefore, the uncertainty generated during a single $e$-fold becomes $(2n+1)(H/2\pi)^2$.
 Of course, it is not necessary that different $\textbf{k}$ mode states share the same  excitation number, then the uncertainty would be smaller than $(2n_{\rm max}+1)(H/2\pi)^2$.
From now on,  we restrict our attention to the simplest case that all the $\textbf{k}$ modes share the same  excitation number to make the discussion simple and emphasize the effects of large $n$ more explicitly.

\section{Condition for eternal inflation}
\label{sec:eternal}

The basic idea underlying eternal inflation is that the inhomogeneous quantum fluctuations contribute to the uncertainty  in the classical inflaton displacement in an accumulative way as they behave like the classical fluctuations after the horizon crossing.
 The probability distribution for the inflaton displacement is obtained by solving the Fokker-Planck equation.
 It describes the random walk of the inflaton trajectory induced by the Gaussian noise, the averaged effect of the quantum fluctuation.
 Then the cosmological perturbation theory tells us that  the  inflaton displacement after $\Delta N_e$ follows the Gaussian probability distribution with the uncertainty given by $(H/2\pi)^2 \Delta N_e$, which is consistent with the $n=0$ solution to the WDW equation. 
 On the other hand, some causal patches in the universe may be in the  higher excitations, $n>0$.
 Indeed, even if the patch we belong to is in the  $n=0$ state, it may be evolved from the patch in the superposition of the $n=0$ state and the higher excitations.
 As a single patch becomes $e^3\simeq 20$  patches after  a single $e$-fold, some of 20 patches can be in the higher excitation, the probability of which is determined by how different excitations are superposed.
 Then we need to investigate whether  the inflaton displacement for the higher excitation patch still follows the Gaussian probability distribution, in which the uncertainty is given by $(2n+1)(H/2\pi)^2$.
 As we will see, the Gaussian distribution is a quite plausible ansatz under the assumption that the classical trajectory of inflaton is regulated by the random walk.
 From this, we  find that the enhanced uncertainty for the higher excitation gives the relaxed condition for eternal inflation compared to that for the $n=0$ case.
 
 \subsection{Probability distribution for the inflaton displacement with $n>0$}
 
 As we have seen, when the causal patch is in the $n$th excitation state,  quantum fluctuation for the $\textbf{k}$ mode has an uncertainty  given by $(n+\frac12)\rho_k^2$.
 This value indeed is a half of the `maximal displacement' squared  in the simple harmonic oscillator analogy.
 For large $n$, the maximal displacement corresponds to the most probable value of $\tilde{\varphi}_\textbf{k}$.
 Meanwhile, the quantum fluctuations contribute to the probability distribution for the classical inflaton displacement  after the  wavelength is stretched  beyond the horizon.
  Then $k$ stands for time when the fluctuation begins to contribute through the relation $k=a(t) H$.
 Hence, we can convert  the uncertainty in the {\it $\textbf{k}$ mode quantum fluctuation} $\tilde{\varphi}_k$ into the uncertainty in the {\it inflaton displacement} $\varphi(t)$ generated per unit $e$-fold, $(2n+1)(H/2\pi)^2 \Delta N_e$.

 The situation above can be modelled in a following way. 
 We first divide the unit $e$-fold into infinitesimal intervals, such that $\Delta N_e= M dN_e$ with $d N_e \to 0$ and $M\to \infty$.
 After $dN_e$, the inflaton moves either forward or backward by the unit length $\sqrt{(2n+1)dN_e}(H/2\pi)$.
 Other values of displacement are assumed to be not so much probable so neglected.
 We can compare this with what actually happens, in which the probability for $\tilde{\varphi}^2_{\textbf k}$ is maximized at $2(n+\frac12)\rho_k^2$ with the strong concentration for large $n$.
 After converting $k$ into time when the mode begins to cross the horizon, we can say $\varphi(t)$ moves by either   $+\sqrt{2(2n+1)dN_e}(H/2\pi)$ or $-\sqrt{2(2n+1)dN_e}(H/2\pi)$ by neglecting the probability for $\tilde{\varphi}_\textbf{k}$ to have other values.
 We note that the uncertainty $(n+\frac12)\rho_k^2$ which gives the unit length and the maximal displacement squared $2(n+\frac12)\rho_k^2$ at which the probability is maximized are different by  the factor 2.
This factor 2 ambiguity can be regarded as a correction from nonzero probability at other values of $\tilde{\varphi}_{\textbf k}$ where the probability is not maximized so we can drop the factor 2.
 
 Then the situation is the same as the simple model for a random walk, in which the particle moves either forward with the probability $p$ or backward with the probability $1-p$ by unit length at each step (see, e.g., chapter 1 of \cite{Reif1965}).
 In our case, $p=1/2$ by symmetry of the probability density  $|\psi|^2$ under $\tilde{\varphi}_{\textbf k}\to -\tilde{\varphi}_{{\textbf k}}$.
 Then the probability that the particle ($\varphi(t)$ in our case) moves $m$ units from the original positions after $M$ steps is given by the binomial distribution,
 \dis{P_M(m)=\frac{M!}{\big(\frac{M+m}{2}\big)! \big(\frac{M-m}{2}\big)!}p^{\big(\frac{M+m}{2}\big)}(1-p)^{\big(\frac{M-m}{2}\big)}=\frac{M!}{\big(\frac{M+m}{2}\big)! \big(\frac{M-m}{2}\big)!}\Big(\frac12\Big)^M,}
 in which the mean value and the uncertainty are given by $\overline{m}=M(p-(1-p))=0$ and $(\Delta m)^2=4Mp(1-p)=M$, respectively.
 As $M\to\infty$, we can regard $m$ as a continuous variable, and the binomial distribution can be replaced by the Gaussian distribution with the same mean value and the uncertainty,
\dis{P(m)=\frac{1}{\sqrt{2\pi Mp(1-p)}}e^{-\frac{(m-M[p-(1-p)])^2}{8Mp(1-p)}},} 
  which is a good approximation for $m \ll M$.
  In terms of the unit length $\sqrt{(2n+1)dN_e}(H/2\pi)$, the normalized probability distribution becomes
  \dis{P(\varphi)=\frac{1}{\sqrt{2\pi}\Delta\varphi}e^{-\frac{(\varphi-\overline{\varphi})^2}{2\Delta\varphi^2}},\label{Eq:Gaussian}  }
  with
  \dis{&\overline{\varphi}=\overline{m}\sqrt{(2n+1)dN_e}\frac{H}{2\pi}=0,
  \\
  &(\Delta\varphi)^2= 4M p(1-p)(2n+1)dN_e\Big(\frac{H}{2\pi}\Big)^2=(2n+1)\Delta N_e \Big(\frac{H}{2\pi}\Big)^2,}
  where  we used $M dN_e=\Delta N_e$ for the last equality.
  
While based on much simplification, the arguments above show that  the Gaussian distribution is a good description for the inflaton displacement for large $n$, up to order one factor on $(\Delta\varphi)^2$ which may come from the factor 2 ambiguity.

\subsection{Eternal inflation in the higher excitations}

Since we have obtained the probability distribution for the  inflaton displacement  generated by quantum fluctuations we can find the condition for eternal inflation by comparing this with the inflaton displacement governed by the classical equation of motion.
For this purpose, we can follow the well-known discussion.
 During a single $e$-fold the inflaton rolls down by $-\dot{\varphi}\Delta t=-(\dot{\varphi}/H)\Delta N_e$.
For eternal inflation to take place, the  inflaton displacement induced by  quantum fluctuations needs to be at least $(\dot{\varphi}/H)\Delta N_e$ to compensate it.
Then the inflaton is pushed up the potential and the vacuum energy does not decrease.
Since a single patch expands to become $e^3\simeq 20$ patches after a single $e$-fold, we require that the vacuum energy does not decrease in at least one of these patches.
Therefore, the condition for eternal inflation is written as, using the Gaussian probability distribution \eqref{Eq:Gaussian}, 
\dis{{\rm Pr}\big(\varphi>\frac{\dot{\varphi}}{H}\big)=\int_{\dot{\varphi}/H}^\infty d\varphi P(\varphi)= \frac12{\rm erfc}\Big(\frac{\dot{\varphi}/H}{\sqrt{(\Delta\varphi)^2}}\Big)>e^{-3},}
i.e., the total probability that the inflaton is pushed up the potential is larger than $e^{-3}$. 
We note that erfc$(x)\sim [e^{-x^2}/(x\sqrt{\pi})](1-\frac{1}{2x^2}+\cdots)$.
While this series expansion is valid for $|x|\gg 1$, it is enough for the rough estimation of the above inequality, which results in $\dot{\varphi}\lesssim [\sqrt3/(2\pi)]\sqrt{2n+1}H^2$.
From \eqref{Eq:epsilon}, this condition is equivalent to 
\dis{\epsilon \lesssim\frac{3}{8\pi^2}(2n+1)\frac{H^2}{\mpl^2},\label{Eq:eternalcond}}
i.e., if the patch in the large $n$ state expands, the condition for  eternal inflation is relaxed by a factor  $2n+1$.
We note that since the WDW equation we solved assumes almost constant $H$, $\epsilon \ll 1$ is required.
Hence, the bound for arbitrary large $n$, say, $n \sim   \mpl^2/H^2$ is meaningless.

 In general, the state of a causal patch is a superposition of different excitations.
 For a concrete discussion, consider the evolution of a single patch in the superposed state $|\psi\rangle=c_0 |n=0\rangle+ c_{10}|n=10\rangle$ with $|c_0|^2+|c_{10}|^2=1$.
  Then a set of $e^3$ patches evolved from this original patch after a single $e$-fold can be regarded as an ensemble of the same quantum state $|\psi\rangle$.
 When the power spectrum and other cosmological parameters are `measured' by  observers scattered in $e^3$ patches, the state in each patch would `collapse' into either $n=0$ or $n=10$, with probabilities $|c_0|^2$ and $|c_{10}|^2$, respectively.
 That is, among $e^3$ patches about $|c_0|^2\times e^3$ patches are in the state $|n=0\rangle$ and about $|c_{10}|^2\times e^3$ patches are in the state $|n=10\rangle$.
 Once the collapse takes place, the state of the patch is no longer superposed and each patch has a different eternal inflation condition for a next $e$-fold : when $n=0$ eternal inflation takes place for $\epsilon \lesssim H^2/\mpl^2$ but when $n\gg 1$, say, $n=10$, even larger  $\epsilon$ satisfying $\epsilon \lesssim 2n (H^2/\mpl^2)=20\times (H^2/\mpl^2)$ allows eternal inflation.

\subsection{Nonlinear interaction effects} 
\label{subsec:nonlinear}
 
 The Hamiltonian \eqref{Eq:Hamkmode} we  considered is quadratic in $\tilde{\varphi}$, in which  nonlinear (cubic or higher order in $\tilde{\varphi}$) interactions have been neglected.
 This is quite a reasonable assumption since the size of nonlinear interaction is suppressed by  $\epsilon$ as well as $H/\mpl$, both of which are much smaller than $1$.
 For example, the cubic interaction typically has the form of \cite{Maldacena:2002vr}
 \dis{{\cal H}_{\rm cubic}\sim a^3 \epsilon^3 \mpl^2 H^2 {\cal R}^3 +\cdots \sim a^3\epsilon^{3/2}\frac{H^2}{\mpl}\tilde{\varphi}^3+\cdots.}
From this, we can estimate  the transition probability between two different excitations.
For this purpose we define the creation/annihilation operators \cite{Lewis:1968tm}
\dis{&A^{R, I}_k=\frac{1}{\sqrt2}\Big[\frac{\tilde{\varphi}^{R, I}_{\textbf k}}{\rho_k}+i(\rho_k \tilde{\pi}^{R, I}_{\textbf k}-M \dot{\rho}_k \tilde{\varphi}^{R, I}_{\textbf k})\Big],
\\
&(A^{R, I}_k)^\dagger=\frac{1}{\sqrt2}\Big[\frac{\tilde{\varphi}^{R, I}_{\textbf k}}{\rho_k}-i(\rho_k \tilde{\pi}^{R, I}_{\textbf k}-M \dot{\rho}_k \tilde{\varphi}^{R, I}_{\textbf k})\Big].}
In terms of these operators,  the Lewis-Riesenfeld invariant for each $\textbf{k}$ mode is written as $I=A^\dagger A+\frac12$ and the wavefunction $\psi_n$ can be regarded as the excitation $|n (t)\rangle=(n!)^{-1/2}(A^\dagger)^n|0\rangle$.
Then we immediately find $\tilde{\varphi}^{R, I}_{\textbf k}=(\rho_k/\sqrt2)[A^{R, I}_k+(A^{R, I}_k)^\dagger]$ hence the cubic term $\tilde{\varphi}^3$ contains various combinations $A^3$, $(A^\dagger)^3$, $A^2 A^\dagger$, and $(A^\dagger)^2 A$ up to ordering, inducing the transition between different excitations.
We also note that  possible  transitions are also determined by the momentum conservation.
Then transitions between states in which various excitation numbers are assigned to each of the $\textbf{k}$ modes take place  but we do not consider it in detail and concentrate on the size of the transition amplitude resulting from  the cubic interaction.
We just treat different $\textbf{k}$s contributing to the transition amplitude as a single $\textbf{k}$ for the rough estimation.

Now consider the wavefunction in the form of 
\dis{|\psi\rangle=\sum_n c_{\{n\}} (t) |\{n(t)\}\rangle.}
Here, the basis state $|\{n(t)\}\rangle$ is the product of the $\textbf{k}$ mode states as can be found in \eqref{Eq:kprod}, and $\{n\}$ denotes a set of excitation numbers assigned to each of the $\textbf{k}$ mode states.
 Indeed, even if   $\textbf{k}$ modes in the initial state share the common excitation number, the  cubic interaction induces the transition to the state in which different excitation numbers are assigned to each of the $\textbf{k}$ mode states.
We also note that the time dependence in the coefficients $c_{\{n\}}(t)$ comes from the cubic interaction. 
Since the total Hamiltonian density is given by ${\cal H}_{\rm tot}={\cal H}+{\cal H}_{\rm cubic}$ where ${\cal H}$ is the quadratic Hamiltonian density we have considered (hence $i\dot{\psi}_n={\cal H}\psi_n$), we find that $c_{\{n\}}(t)$ with the initial condition $c_{\{n\}}(0)=\delta_{ni}$ for some initial excitation $|\{i\}\rangle$ is given by
\dis{c_{\{n\}}(t)\simeq \delta_{ni}-i\int d^4x \langle \{n\}(t)|{\cal H}_{\rm cubic}|\{i\}(t)\rangle.}
If our initial time is much larger than $(1/H)\log(k/H)$, $\rho_k\simeq H/k^{3/2}$ is almost constant and the Lewis phase,
\dis{\alpha_n=-\big(n+\frac12\big)\int^t dt \frac{dt'}{a^3\rho_k^2}\simeq -\big(n+\frac12\big)\frac{k^3}{(a(t)H)^3}}
 is the only phase in the wavefunction $\psi_{\textbf{k}, n}$.
Then the transition amplitude from $|\{i\}(t)\rangle$ to $|\{n\}(t)\rangle$ per unit volume is given by (up to $n$ dependence)
\footnote{Of course, the rigorous calculation is done in the momentum space and integration over $x$ gives the overall $\delta$ function which is equivalent to the overall volume. 
We further multiply the result by $e^{-3Ht}$ for the transition amplitude per `unit volume' as the volume expands as $e^{3Ht}$.}
\dis{\int d t \langle \{n\}(t)|{\cal H}_{\rm cubic}|\{i\}(t)\rangle &\sim e^{-3Ht}\int dt e^{-i(\alpha_n-\alpha_i)}\Big(\frac{\rho_k^3}{2^{3/2}} e^{3 Ht} \epsilon^{3/2}\frac{H^2}{\mpl}\Big)
\\
&\sim  \frac{H^3 }{k^{9/2}}\epsilon^{3/2}\Big(\frac{H}{\mpl}\Big)e^{-i\frac{k^3}{(a_0 H)^3}e^{-3 Ht}(n-i)} ,\label{Eq:transamp}}
 where we keep the dominant term at large $t$ only.
 As expected, the probability for the transition between different excitations is suppressed by small $\epsilon$ and $H/\mpl$.
While the transition amplitude typically enhanced by the large excitation number, an arbitrary large value of it invalidates our perturbative estimation.

  From this, we can find a time scale after which the initial excitation annihilates into the lower excitations like the ground state ($n=0$).
  Estimation in \eqref{Eq:transamp} indicates that the transition amplitude for  the Hubble patch of size $H^{-1}$ over the time interval $\Delta t$ is roughly given by
  \dis{\frac{e^{3 H\Delta t}}{H^3}\times \frac{H^3}{k^{9/2}}\epsilon^{3/2}\Big(\frac{H}{\mpl}\Big),}
  where we ignored the phase as it does not contribute to the probability after being squared.
  We also note that since a scale factor has a dimension $-1$ in our discussion so far, the spatial coordinate hence the wavenumber $k$ is dimensionless. 
  As the initial scale factor is given by $H^{-1}$, the wavenumber $k$ in \eqref{Eq:transamp} is understood as $k/H$ in which $k$ has a dimension $1$.
  Then the transition probability becomes ${\cal O}(1)$ when
  \dis{\Delta t \sim \frac{1}{H}\log\Big[\frac{1}{\epsilon^{3/2}}\frac{\mpl}{H}\Big(\frac{k}{H}\Big)^{9/2}\Big].}
  Therefore, as long as the argument in the logarithm is much larger than ${\cal O}(1)$, $\Delta t$ is  longer than $H^{-1}$, which implies that the transition does not play a crucial role during a single $e$-fold.
  This indeed is supported by small values of $\epsilon, (H/\mpl) \ll 1$.
  We note that in eternal inflation the superhorizon modes that cross the horizon during $\Delta t$ are relevant. 
  If we consider eternal inflation at the early stage of inflation, the long wavelength modes satisfying $k < H \epsilon^{1/3}(H/\mpl)^{2/9}$ cross the horizon during $\Delta t$ which  may allow $\Delta t <H^{-1}$.
 However, such a small value of $k$  spoils the perturbative expansion which is implicit in \eqref{Eq:transamp}.
 Moreover, the factor $k^{3/2}$ comes from the estimation of $\rho_k$ in the large $t$ limit (see \eqref{Eq:rhok}).
 Hence, by restricting our attention to eternal inflation at sufficiently late time such an extreme superhorizon modes are irrelevant and $\Delta t$ is typically larger than $H^{-1}$.

 On the other hand, the nonlinear interaction allows the interaction between long- and short wavelength modes, which is not present in the quadratic action.
 Especially, superhorizon modes lose their quantum nature  through the interaction with the subhorizon modes.
 Such `decoherence' provides the mechanism for the quantum-to-classical transition.
  But decoherence does not occur immediately after the  horizon   crossing : it takes the `decoherence time' given by  $t_{\rm dec} \sim (1/H)\log[(\sqrt{\epsilon}\mpl)/((\epsilon+\eta)H)]$ in addition \cite{Nelson:2016kjm}. 
 \footnote{ For the tensor perturbation, the decoherence time is given by $(1/H)\log(\mpl/H)$ \cite{Gong:2019yyz}.
 The fact that this time scale has to do with the time scale for the trnas-Planckian mode to escape the horizon leads to the `trans-Planckian censorship conjecture'  \cite{Bedroya:2019snp} (see also \cite{Hayden:2007cs, Sekino:2008he} for the black hole analogy).
 We also note that the decoherence does not mean the complete disappearance of the quantum effect \cite{Gong:2020gdb}.}
 Since eternal inflation is the effect of the superhorizon modes, we expect that the delay in becoming classical fluctuations by decoherence changes the probability distribution of the inflaton displacement.
 More concretely, the $\textbf{k}$ mode becomes classical not at $t=(1/H)\log(k/H)$ satisfying $k=aH$, but $(1/H)\log(k/H)+t_{\rm dec}$.
 From this, \cite{Boddy:2016zkn} suggested that the mode which crosses the horizon at $t_i-t_{\rm dec}$, rather than   $t_i$ contributes to the probability distribution for the inflaton displacement generated at $t_i$.
 It means that the quantum fluctuation which becomes `completely' (after decoherence) classical  at $t_i$ affects  the inflaton displacement which rolls down from $\varphi(t_i)+\dot{\varphi}t_{\rm dec}$, rather than $\varphi(t_i)$. 
 This in fact does not cause a significant change since we just need to shift the time $t$ by $t_{\rm dec}$.
 The uncertainty   does not change, and the classical inflaton displacement during a single $e$-fold also does not change at the formal level : $\Delta \varphi=[\varphi(t_i)-\dot{\varphi}(H^{-1}-t_{\rm dec})]- [\varphi(t_i)+\dot{\varphi}t_{\rm dec}]$, which is just $-\dot{\varphi}H^{-1}$.
 The leading decoherence effect appears in the shifted time at which $\dot{\varphi}$ is calculated \cite{Boddy:2016zkn}.
 For example, as suggested in \cite{Boddy:2016zkn}, we can take $\dot{\varphi}$ to be the value  when $\varphi$ passes the averaged position $\frac12[[\varphi(t_i)-\dot{\varphi}(H^{-1}-t_{\rm dec})]+ [\varphi(t_i)+\dot{\varphi}t_{\rm dec}]]=\varphi(t_i)+\dot{\varphi}t_{\rm dec}-\frac12 \dot{\varphi}H^{-1}$.

 \subsection{Entropy consideration}
 \label{subsec:entropy}
 
 As shown in section \ref{subsec:nonlinear}, the nonlinear interaction can induce the transition between different excitations even though the probability is suppressed by small $\epsilon$ and $H/\mpl$.
 Then we can ask if there is a `direction' of the transition toward some specific $n$.
 Indeed,  in the adiabatic approximation, the $n=0$ state is regarded as a ground state, so we expect that after a long enough time, transitions  lead to the settlement of the patches in the $n=0$ state.
 While the exponential expansion in time invalidates the adiabatic approximation in our case, we may reach the same conclusion by considering the second law of thermodynamics along the line of discussion in \cite{Bousso:2006ge}.

 For this purpose, we compare the changes in the entropy during $\Delta N_e$ originated from two cases : one from the classical displacement $\Delta_{\rm cl}\varphi=-(\dot{\varphi}/H)\Delta N_e$ and another from the quantum fluctuations $\Delta_{\rm qu}\varphi=\sqrt{(2n+1)dN_e}(H/2\pi)$.
 Since the entropy of dS spacetime is given by $(\mpl^2/4)\times$(area of the horizon)$=\pi \mpl^2/H^2$ \cite{Gibbons:1977mu},  it is increased by the classical trajectory of the inflaton which decreases $H$ but decreased by  the quantum fluctuations contributing to eternal inflation which increase $H$.
 To see this explicitly, we first estimate the increment of the entropy induced by the classical slow-roll of the inflaton :
 \dis{\Delta_{\rm cl} S=-2\pi\mpl^2\frac{\dot H}{H^4}\Delta N_e=2\epsilon S \Delta N_e.} 
 To see the physical meaning of it, we note that the size of the density perturbation generated during a single $e$-fold   is given by \cite{ArkaniHamed:2007ky}
 \dis{\frac{\Delta \rho}{\rho} \sim H\Delta_{\rm qu}t =\frac{H}{\dot{\varphi}}\Delta_{\rm qu}\varphi=\frac{\sqrt{2n+1}}{2\pi \sqrt{2\epsilon}}\frac{H}{\mpl},}
 in terms of which the classical increment of the entropy during a single $e$-fold is written as $\Delta_{\rm cl}S \simeq(2n+1)(\rho/\Delta\rho)^2$ up to order one coefficient.
 If the condition for eternal inflation \eqref{Eq:eternalcond} is satisfied, $\Delta_{\rm cl} S \lesssim (2n+1)$ (for $\Delta N_e=1$) hence the density perturbation $\Delta\rho/\rho$ becomes larger than $1$, i.e., large enough to generate the primordial black hole and the perturbative approach becomes unreliable.
 We note that the enhancement factor $2n+1$ is irrelevant to the bound on the density perturbation.
 
 On the other hand, the decrement of the entropy induced by the quantum fluctuation is estimated as
 \dis{\Delta_{\rm qu} S&=\frac{d S}{dt}\Delta_{\rm qu}t=-2\pi\frac{\mpl^2}{H^3}\frac{\dot H}{\dot \varphi} \Delta_{\rm qu}\varphi= -\frac{\sqrt{\epsilon}}{\sqrt2}\sqrt{(2n+1)\Delta N_e}\frac{\mpl}{H}
 \\
 &=-\frac12\sqrt{2n+1}(\Delta_{\rm cl}S)^{1/2},}
 where we used $\epsilon=+\dot{H}/H^2$ since $\dot{H}>0$ in eternal inflation.
 This shows that 
  the decrease in the entropy by the eternally inflating patch gets larger as $n$ gets larger, so it is difficult to overcome it by the increase in the entropy by the slow-rolling patch.
 Then we  can infer that the increase in the total entropy as required by the second law of thermodynamics prefers the $n=0$ state, rather than the large $n$ state.
 Of course, it will take much  time  until the states of patches are stabilized to $n=0$ state, hence eternal inflation induced by large $n$ excitations is not ignorable.

 \section{Conclusions}
\label{sec:conclusion}

 In (quasi)dS spacetime, the quantum fluctuations of the curvature perturbation, the gauge invariant quantum fluctuation of the inflaton evolve into the classical ones as they cross the horizon, generating the uncertainty in the classical inflaton displacement   accumulatively. 
 When the Hubble parameter $H$ varies much slowly such that $\epsilon$  is very tiny, the probability that the vacuum energy does not decrease is not negligible then the  eternal inflation takes place. 
 In the cosmological perturbation theory, the power spectrum predicts that the uncertainty is given by $(H/2\pi)^2\Delta N_e$.
 On the other hand, the probability distribution for the curvature perturbation can be obtained by solving the WDW equation.
 This shows that the uncertainty obtained from the power spectrum in cosmological perturbation theory is one of various possibilities corresponding to the `ground state' of the time dependent harmonic oscillator.
 Moreover, the WDW equation also provides the larger uncertainty $(2n+1)(H/2\pi)^2\Delta N_e$ in which  the integer $n$ labels a tower of `excitations'.
 As a result, the bound on the slow-roll parameter $\epsilon$ that allows  eternal inflation is given by $\epsilon \lesssim (2n+1)(H/\mpl)^2$.
 The value of  $\epsilon$ much larger than the well-known bound $(H/\mpl)^2$ can give rise to  eternal inflation if some of causal patches are in the $n\gg 1$ state.
  This can be one of examples that the WDW equation  captures what we may have missed in the perturbative approach.
 On the other hand, nonlinear interaction induces the transition between different excitations, and the second law of thermodynamics seems to prefer the $n=0$ state.
  Even in this case, the transition probability is small, hence the existence of higher excitations during long enough period results in the evolution of the universe different from that of the universe in a mere $n=0$ state as the bound for eternal inflation is relaxed.

\subsection*{Acknowledgements}

MS is grateful to Jinn-Ouk Gong and Gary Shiu for discussion and comments    while this work was under progress.
%

%


\appendix

\renewcommand{\theequation}{\Alph{section}.\arabic{equation}}

\section{Uncertainty for the infrared modes}
\label{app:IRuncert}
\setcounter{equation}{0}

In the discussion on eternal inflation, we are interested in the accumulated uncertainty in the inflaton displacement during $\Delta N_e=1$.
Since it considers the change in uncertainty, the common contribution from the infrared mode is irrelevant.
On the other hand,  an uncertainty from the infrared mode $k \ll a |m|$ is formally equivalent to the probability distribution for the  massive, homogeneous $\partial_i\varphi=0$ scalar field which is not affected by the super-horizon mode fluctuations as studied in section \ref{subsec:WDWSolcal}.
Since it is instructive to investigate the time change in the uncertainty, we discuss it in this appendix.

The differential equation for $\rho$ in this case is,
  \dis{\ddot{\rho}+3H\dot{\rho}+m^2\rho=\frac{e^{-6 H t}}{a_0^6 \rho^3},}
  the solution to which is given by
 \dis{&\rho=\frac{H}{2}e^{-\frac32 H t}\frac{\Big[ e^{-3H t \sqrt{1-\frac49 \frac{m^2}{H^2}}} + 16 e^{3Ht\sqrt{1-\frac49 \frac{m^2}{H^2}}}\Big]^{1/2}}{(a_0H)^{3/2} \big[9-4 \frac{m^2}{H^2}\big]^{1/4}},\quad\quad {\rm or}
 \\
 &\rho=\frac{H}{2}e^{-\frac32 H t}\frac{\Big[16 e^{-3H t \sqrt{1-\frac49 \frac{m^2}{H^2}}} +  e^{3Ht\sqrt{1-\frac49 \frac{m^2}{H^2}}}\Big]^{1/2}}{(a_0H)^{3/2} \big[9-4 \frac{m^2}{H^2}\big]^{1/4}}. }
 In any case, as time goes on, $\rho$ behaves like
 \dis{\rho \sim \frac{H}{(a_0 H)^{3/2}}e^{-\frac32 H t+\frac32 Ht \sqrt{1-\frac49 \frac{m^2}{H^2}}},\label{Eq:homsol}}
 up to order one factor.
 If we are interested in the evolution of a single causal patch, we can take the value of $a(t)$ at $t=0$ to be $H^{-1}$, or $a_0=H^{-1}$.
 Then since $m^2 \ll H^2$, we have $\rho \sim H$ like the massless case.
 
 We also note that the exponent in \eqref{Eq:homsol} is expanded as $-\frac13 H(m^2/H^2)t$.
 When $m^2<0$, the uncertainty gradually increases in time, reflecting the instability of the classical trajectory under the quantum fluctuation.
 The uncertainty reaches the cutoff scale, say, Planck scale $\mpl$ after $t=3/(H|\eta|)\log(\mpl/H)$.
 Here, we estimate $m^2=\eta H^2$ for the curvature perturbation, in which $\eta$ is identified with $\mpl^2 V''/V$ in the slow-roll approximation.
 After this time, the quantum fluctuation can disperse to trans-Planckian scale, and we expect that the quantum gravity effect that we are not aware of may emerge.
 For example, the trans-Planckian inflaton displacement by the quantum fluctuation   may make  some particles whose masses are connected to the inflaton field value  descend from UV to spoil the effective theory we have used.
 This is the prediction of the distance conjecture \cite{Ooguri:2006in}, motivated by the descent of KK mode for the large radion value in the presence of the extra-dimension.
  The time scale for the breakdown of effective theory predicted by the distance conjecture, with the help of the Bousso's entropy bound \cite{Bousso:1999xy} is given by $1/(\sqrt{\epsilon_H}H)\log(\mpl/H)$, shorter than our result by a factor of the square root of the slow-roll parameter \cite{Seo:2019wsh, Cai:2019dzj}.
 On the other hand, when $m^2>0$, the uncertainty becomes smaller as time goes on : after $t =3/(\eta H)$,  the uncertainty becomes much smaller than $H$, and the infrared mode more or less follows the classical trajectory.

\end{document}